\documentclass[notitlepage,12pt]{report}

\newcommand{\be}{\begin{equation}}
\newcommand{\ee}{\end{equation}}
 
\newcommand{\bea}{\begin{eqnarray}}
\newcommand{\eea}{\end{eqnarray}}

\usepackage{csquotes}

\begin{document}

\centerline{\bf
The Early Scientific Contributions of J. Robert Oppenheimer}


\centerline{\it
Why did the scientific community miss the black hole opportunity?}

\hfil




{\footnotesize 

\centerline{
M.~Ortega-Rodr\1guez, H.~Sol\1s-S\'anchez, 
E.~Boza-Oviedo,
K.~Chaves-Cruz,}
\centerline{ 
M.~Guevara-Bertsch,
M.~Quir\'os-Rojas, 
S.~Vargas-Hern\'andez, and}
\centerline{A.~Venegas-Li}

\centerline{
Escuela de F\1sica, 
Universidad de Costa Rica, 11501-2060 San Jos\'e, Costa Rica} }


\hfil

\begin{abstract}
\noindent
We aim to 
carry out an assessment of the scientific value of Oppenheimer's
research on black holes 
in order to determine and weigh possible factors 
to explain its neglect by the scientific 
community, and even by Oppenheimer himself.
Dealing primarily with the science
and looking closely at the scientific culture and 
the scientific conceptual belief system of the 1930s, 
the present article
seeks to supplement the existent literature 
on the subject by enriching the explanations 
and possibly complicating the guiding questions.
We suggest a rereading of Oppenheimer as a more intriguing, ahead-of-his-time 
figure.\footnote{We
have greatly benefited from discussions with Barton Bernstein,
which led to the organization of a multidisciplinary conversation
with historians, philosophers, and physicists, among others, at Stanford
University's 
Hansen Experimental Physics Laboratory on January 31, 2014,
with a follow-up on January 30, 2015 in the History Dept. These Stanford
sessions were themselves continuations 
of earlier conversations at Universidad de Costa Rica.
The insightful and thorough accompanying article by Barton Bernstein 
(which has the merit that does not shy away from the science) 
and this paper complement each other and are best read together.
To the date March 12, 2017, Barton Bernstein's paper has not been
published yet.}
\end{abstract}

\hfil

\noindent{\bf Background}

The 1930s witnessed a tremendous growth in our understanding 
of stars.
Not only did 
Hans Bethe 
and others solve the long-standing problem
of stellar energy production by means of nuclear fusion, 
but the recently discovered neutron (1932) allowed
for speculation about the existence of more extreme physics.
In this way, Fritz Zwicky and Lev Landau considered the possibility of
stars composed entirely of neutrons.
Along similar lines, J.~Robert Oppenheimer 
became deeply interested in the problem of stellar stability,
leading to an acute interest in {\it total} stellar collapse.
Oppenheimer invented the concept of black holes.

Apparently, though, Oppenheimer's move was too extreme.
Despite the fact 
that these ideas are considered milestones today,
in 1939 and 
for reasons that are not completely understood
they fell into oblivion for two decades,
failing to capture the attention of most physicists
(Landau being a notable exception).

Freeman Dyson calls Oppenheimer's black hole work his ``only revolutionary 
contribution to science.''
Furthermore, Dyson considers 
``the outstanding mystery in Oppenheimer's life''
the fact that even Oppenheimer failed
to grasp the importance of his own discovery.\footnote{Dyson, F. (2013, August 15),
{\it Oppenheimer: The shape of genius},
retrieved from
http://www.nybooks.com/articles/archives/2013/aug/15/oppenheimer-shape-genius/\label{dyson}}

Indeed, 
Oppenheimer never regained interest in the topic,
a potential Nobel winner.
When biographer Abraham Pais asked him what his most 
important contribution to science had been, he referred to his
electron/positron work, not a word on 
astrophysics.\footnote{Pais, A. (2006), {\it J.~Robert Oppenheimer, a life},
New York: Oxford University Press, 33.}


\hfil

\noindent{\bf Main Question: Why Did the Scientific Community Miss
the Black Hole Opportunity?}

The present multidisciplinary collaboration
aims to 
carry out an assessment of the scientific value of Oppenheimer's
research on black holes 
in order to determine and weigh possible factors 
to explain its neglect by the scientific 
community, and even by Oppenheimer himself.\footnote{This 
version of the paper can be read without having
technical knowledge of general relativity.}

Not that there is a lack of easy ways to dismiss,
or to address, 
this question.
For example, by arguing that Oppenheimer's discovery was beyond
experimental/observational corroboration and thus
scientifically uninteresting.
But that answer ignores the fact that physics seems many times not to 
care about this circumstance, and that {\it theoretical}
corroborations and elaborations were doable in the 1930s even when the
observational ones were not feasible.

We believe, then, that much insight can be gained from
plunging into the question of the present paper's subtitle.
This subtitle question hints at another question, namely:
What would it have taken for the black hole concept to become an
active field of research in 1939?


This article differs from previous treatments of the subject
in its emphasis. This essay $a)$ deals primarily with the science,
$b)$ attempts to be 
situated in time (i.e., forgetting what came after 1939),
and $c)$ looks closely at the scientific culture and 
conceptual belief system of the 1930s, in particular
it considers
what ``good science'' meant back then.

This paper therefore complements studies using other perspectives, 
such as
career choices, network analyses
(including very counterproductive enmities),
German-Jew frustrated liberal idealism,
the whole ``bag'' of personality traits 
(Oppenheimer's peculiar intellectual impatient style, 
his ``pathological'' interest in everything,
his constant desire to be at the center of things, 
his fierce independence and
{\it Sitzfleisch}\footnote{According to Dyson (ref.~\ref{dyson}), 19,
and literally meaning ``sit still,'' this term refers
to Oppenheimer's inability to sit still and work quietly to finish
a difficult calculation.} problem),
in addition to 
purely contingent factors: war, anti-Semitism, nationality issues, etc.

\hfil


\noindent{\bf Review of the Literature} 


In addition to the well-known biographies of Oppenheimer, 
the subject of the contextualized stellar science of Oppenheimer 
has been touched upon by several authors with 
different backgrounds.

The most detailed account, to our knowledge, is the one given
by historian-of-science Karl Hufbauer,\footnote{Hufbauer, K. (2005),
J.~Robert Oppenheimer's path to black holes,
in C. Carson \& D. A. Hollinger (Eds.),
{\it Reappraising Oppenheimer, Centennial Studies and Reflections} 
(pp.~31--47),
Berkeley:
University of California, Berkeley.\label{hufbauer}} which constitutes our
starting point (see next section).
In addition,
the black hole science is briefly commented on 
books by physicist/journalist 
Jeremy Bernstein\footnote{Bernstein, J.~(2004),
{\it Oppenheimer: Portrait of an enigma},
Chicago:
Ivan R.~Dee.
} and
historian-of-science  
David Cassidy.\footnote{Cassidy, D.~(2005), \label{cassidy}
{\it J.~Robert Oppenheimer and the American century},
New York: Pi Press. 
} 

Kip Thorne, an astrophysicist, presents a comprehensive view of the
circumstances surrounding the black hole conception,
including Oppenheimer's confrontation with theoretical physicist
John Wheeler in 1958
in Brussels.\footnote{Thorne, K.~(1994), 
{\it Black holes and time warps: Einstein's outrageous legacy},
New York: Norton, 209\label{thorne}; details of the confrontation 
can be found in  
Israel, W. (1987),
Dark stars: the evolution of an idea,
in S. Hawking \& W. Israel (Eds.),
{\it 300 Years of Gravitation} 
(pp.~199--276), 
Cambridge:
Cambridge University Press, 229.\label{300y}}
Thorne's account is the most complete from a scientific point
of view with the caveat of being seen through 
modern eyes.

Finally, we must mention
Freeman Dyson's lucid review\footnote{Dyson (ref.~\ref{dyson}).} 
of biographer Ray Monk's 
book on Oppenheimer.\footnote{Monk, R.~(2013),
{\it Robert Oppenheimer: His life and mind}, New York: Doubleday.} 
Dyson actually reviews, albeit briefly, 
the whole Oppenheimer's science debate.


\hfil

\noindent{\bf Summary of Hufbauer's Article}

The article by Hufbauer represents, to our knowledge, 
the most comprehensive
historical study of the black hole quest by Oppenheimer.


Hufbauer carefully explicates the path of
events that led to the publication of the 
three relevant papers
(Oppenheimer \& Serber 1938,
Oppenheimer \& Volkoff 1939,
Oppenheimer \& Snyder 1939\footnote{Oppenheimer, J.~R., 
\& Serber, R. \label{oppserber} (Oct 1, 1938), 
``On the stability of stellar neutron cores,''
{\it Physical Review, 54}, 540;
Oppenheimer, J.~R., \& Volkoff, G.~M. (Feb 15, 1939),
``On massive neutron cores,''
{\it Physical Review, 55}, 374--381; and
Oppenheimer, J.~R., \& Snyder, H. (Sept 1, 1939),
``On continued gravitational contraction,''
{\it Physical Review, 56}, 455--459.}),
including how Oppenheimer became interested
as early as 1933 in high-density stellar physics.
This was facilitated by his simultaneous
interest and competence in
both particle physics and
astronomy, a rather American trait.
(For the benefit of readers not familiar with the papers, 
there is a brief description of each in Appendix A.)

In addition, Hufbauer describes Oppenheimer's efficient use
of available resources, 
including talking to prominent
figures like his Caltech colleague Richard Tolman.
Hufbauer also describes the way in which 
Bethe ``scooped'' Oppenheimer on the 
topic of stellar energy.
Hufbauer then discusses the main results of the Oppenheimer \& Snyder paper:
not only the surprising collapse, but also
how time freezes at the Schwarzschild radius. 

Finally, Hufbauer offers five reasons for the early neglect of
Oppenheimer's papers, in the form of a contrast with
Bethe's more successful experience. Unlike Bethe's
research on stars, Oppenheimer
$a)$ was not addressing a well-defined problem with a large
   following;
$b)$ had no data and was invoking the little-used theory of 
   general relativity;
$c)$ offered a solution that was completely counterintuitive;
$d)$ did not reach out to potential audiences;
$e)$~published his paper just as the war began.



\newpage

\noindent{\bf The Value of Oppenheimer's Work}  

There is no real doubt that Oppenheimer's work on black holes is
considered good science according to our {\it modern} point of view.
The internal logic of the decade-long
development of the ideas about denser and denser astrophysical entities
is very clearly expounded in
Thorne's book using nontechnical language.\footnote{Thorne 
(ref.~\ref{thorne}), 187--197 and 209--219.}
In addition, the citation record of the paper by Oppenheimer \& Snyder
shows a clear {\it delayed} recognition of their ideas, 
in the 1960s. 

The real question is whether Oppenheimer's work was
considered good science according to the standards of the time.
A second, related question is whether he was preeminent or not
among scientists along this line of research.
 
The following quote, taken
from a long, authoritative 
(``a bible in the field'') stellar evolution review from 1962
(and therefore written more than two decades after Oppenheimer's work)
is helpful in this respect:\footnote{Hayashi, C., Hoshi, R., \& Sugimoto, D.
(1962), ``Evolution of the Stars,'' {\it Progr. Theoret. Phys. Supp., 22}, 
95.  See also footnote \ref{bible}.}


\begin{displayquote}
A 
possibility of stellar evolution leading to these extremely dense configurations [more dense than neutron stars] may not be denied, but it will be highly more probable that, before the star reaches such a configuration, its mass will be reduced below its Chandrasekhar limit by mass ejection from its surface, due to an increase in the centrifugal force in the course of contraction.
\end{displayquote}

In the sentence immediately preceding this quote in
the review, 
the work of Oppenheimer
with Snyder and Volkoff is referenced, but not 
Albert Einstein's related article\footnote{Einstein, A.~(1939), \label{einstein} 
``On a stationary system with spherical symmetry
consisting of many gravitating masses,''
{\it Annals of Mathematics, 40}, 922--936.}
(described in Appendix A), 
or any other author's.
This shows that, 
{\it even when the scientific community as a whole
(if we take this comment as representative) 
still did not believe in black holes}, Oppenheimer's work was considered 
authoritative.  The authors could have 
easily been more dismissive of Oppenheimer, who 
had after all disappeared from the field of (what we would now call) 
astrophysics.\footnote{By 1962, Chushiro Hayashi was already 42 years old
and a prestigious
scholar who had received a Professor appointment at Kyoto University
five years before. This strongly reduces the probability of him having paid lip service
to Oppenheimer. 
According to the American Astronomical Society, 
Hayashi's review with Hoshi and Sugimoto 
was considered \label{bible} ``...a bible in the field of stellar 
evolution for a long time, and may be so still.''
See https://aas.org/obituaries/chushiro-hayashi-1920-2010.
}

Further proof of Oppenheimer being considered as a preeminent scientist is the
fact that Landau allegedly included the Oppenheimer \& Snyder paper in 
his ``Golden List'' of classic papers in 
1939.\footnote{Explained in 
Hufbauer (ref.~\ref{hufbauer}), 46 and footnote 77;
Thorne (ref.~\ref{thorne}), 219.}
What we do know for certain is that
Landau and Lifshitz cite the work of Oppenheimer with Snyder 
in their widely read 1951 (Russian) edition of
{\it Statistical 
Physics}.\footnote{Landau, L., \& Lifshitz, E. (1951),
{\it Statisticheskaya Fizika}, Moscow: Fizmatgiz.}
(The corresponding English edition\footnote{Landau, L., \& Lifshitz, E. (1958),
{\it  Statistical Physics}, Oxford: Pergamon.}
came 
out in 1958, and constitutes to the best of our knowledge the first
critical citation of Oppenheimer's black hole work in the Western 
World.\footnote{There are a few earlier citations of the work 
of Oppenheimer with Snyder, but these are made in passing
and refer not to star collapse but to more normal stellar dynamics. 
See
Johnson, M. (1946), ``Atomic possibilities underlying stellar catastrophe,'' 
{\it The Observatory, 66}, 248--254;
Borst, L. B. (1950), ``Supernovae,'' {\it Physical Review, 78}, 807--808;
and
Vaidya, P. C. (1951), ``Nonstatic solutions of Einstein's field equations for spheres of fluids radiating energy,'' {\it Physical Review, 83}, 10--17.})

In this Russian book, 
Landau and Lifshitz fully support
the relevant ideas: 
\begin{displayquote}
Such a study [the one by Oppenheimer and Snyder] has been carried out only for the simplest
case of the equation of state $P=0$, i.e. for a sphere consisting of a very
thin substance; {\it it probably gives also a correct indication of the nature of 
the process for the general case} of an exact equation of state
[emphasis added].
\end{displayquote}

One should also point out that
there is not a single published attack 
to Oppenheimer's ideas on black holes until the
publication of a paper by Tullio Regge and Wheeler 
in 1957\footnote{Regge, T., \& Wheeler, J.~A.~(1957), ``Stability of a 
Schwarzschild singularity,'' {\it Physical Review, 108}, 1063--1069.} 
(see immediately below), in which the attack is tacit as 
Oppenheimer is not referenced.
The only piece resembling an attack on black holes 
before 1957 was Einstein's 1939 paper,
but this article faded away quickly.  It was not cited
until 1953, and then only to be torn apart by 
Amalkumar Raychaudhuri.\footnote{Raychaudhuri, A.~(1953),
``Arbitrary concentrations of matter and the Schwarzschild singularity,''
{\it Physical Review, 89}, 417--421.} 

The fact that Oppenheimer's ideas survived Einstein's assault
is significant.
Also significant is the fact that no other scientist
published anything else on the subject of black holes
until the late 1950s.
A careful scrutiny of the articles written by 
Landau, Zwicky, Bethe, Richard Tolman, George Gamow, 
Robert Serber, George Volkoff, and Hartland Snyder
shows nothing on this. 


The first paper dealing with the subject is the one by 
Regge and Wheeler in 1957 mentioned above,
where the authors proposed wormholes as a way to avoid total collapse of the star.
It is important to stress that even though this 1957 paper does not reference 
Oppenheimer explicitly, it is clear that the paper is presented as
a criticism of Oppenheimer's ideas on indefinite contraction: 
there is a bold and unnecessary emphasis on the concept of ``stability''
all throughout the paper, 
including the first word in the title and the last sentence of the paper.
A casual reader could be thus forgiven for thinking that
the paper is not so much about discussing wormhole physics
as being a defense of stellar stability under extreme conditions.

David Finkelstein wrote a paper in 1958 where, though not directly
addressing total collapse, he established the Schwarzschild radius as
a surface of 
``no return.''\footnote{Finkelstein, D.~(1958), ``Past-future asymmetry of the gravitational field of a point particle,'' {\it Physical Review, 110}, 965--967.}
Finkelstein did not reference Oppenheimer either.

In 1960, Wheeler wrote a paper on behalf of Martin Kruskal in 
which black holes are finally 
acknowledged.\footnote{Kruskal, M.~(1960),
``Maximal extension of Schwarzschild metric,'' {\it Physical Review, 119}, 
1743--1745.} 
It is significant that
Wheeler is not listed as co-author of 
the paper even though he did the 
actual 
writing,\footnote{Wheeler, J. A., \& Ford, K. (2000), 
{\it Geons, Black Holes \& Quantum Foam: A Life in Physics}, 
New York: Norton, 745.}
and that Oppenheimer went unreferenced one more
time.


In addition to these publications, there is unpublished 
work of Wheeler and (independently) Yakov Zel'dovich 
in the late 1950s, using computers, as reported by Thorne.\footnote{Thorne 
(ref.~\ref{thorne}), 197 and 240.}
One must also not forget about the 1958 Brussels confrontation of Wheeler with Oppenheimer
mentioned above.



\hfil

\noindent{\bf Four Arguments}

We now plunge into the question in the subtitle of this paper:
Why did the scientific community miss the black hole opportunity?
We note that,
even though the five reasons listed by Hufbauer
are sensible and generally agreed upon,
we believe that they could benefit from being 
elaborated (as in the ``too esoteric'' and ``not earned the right'' arguments below)
and extended (as in the ``wrong episteme'' and ``wrong relativistic ontology''    arguments).

The last two arguments are of a Kuhnian, ``history of ideas'' flavor and 
are offered here to complement more conventional approaches.
Even if somewhat Foucauldian, they try to provide 
a fresh perspective on how the conceptual framework 
of knowing and discovery 
could have been very different back then. 

Before starting, we
discuss some general considerations (in the next section)
and make the perhaps unnecessary proviso
that the four arguments below are not independent among themselves
nor with extra-scientific factors.\footnote{Part of 
the conundrum's answer is clearly extra-scientific.
To give but one example, 
take Oppenheimer and (Caltech colleague) Zwicky's
refusal even to acknowledge each other's papers.
Oppenheimer never used the word 
``neutron star.'' See Thorne (ref.~\ref{thorne}), 206.}

\hfil

\noindent{\bf General Considerations}

\begin{flushright} 
{\footnotesize 
{\sl The past is a foreign country: they do things differently there.} \\
L. P. Hartley}
\end{flushright}

For a trained scientist, 
the main difficulty in a project like this one is 
effectively situating oneself in the 
conceptual framework of the time,
forgetting what came after 1939.

For us,
a black hole is an exciting opportunity. Back then, it was 
a nuisance in need of quick repair. 
Just to begin, think how
a physicist living and working in the 1930s would
have perceived an intellectual world very different from ours:

In the first place, the disciplinary landscape
was very different. There was no ``solid-state'' discipline,
and there were no ``astrophysics'' or ``cosmology'' disciplines
in the institutional sense, in sharp contrast
with the prestigious particle and nuclear physics disciplines.
This lack of disciplinary affiliation made it difficult for 
a person like Gamow, even as late as the 1950s,
to find a scientific audience for his cosmology ideas. 
On top of everything, the United States was not a scientific
power like it is today.\footnote{A general reference for the 
statements made in this section of the paper is
Kragh, H.~(2002), \label{kragh} {\it Quantum generations: A history of physics
in the twentieth century},
Princeton, NJ: Princeton University Press.}






In the second place,
most of the related basic knowledge we take for granted today was 
absent:
The mechanism for stellar energy was unknown,
only being teased out in 1938 by Bethe and Carl von Weizs\"acker. 
The neutron was a new thing, and the muon did not appear
until 1937.
Astronomers
had not quite finished digesting the fact that galaxies
were not nebulas in the Milky Way.\footnote{We are
grateful to Prof.~James Bjorken (Stanford) for his comments on this 
particular issue and for his interest in this paper's discussion.
He recalls how, as late as 1950, the multi-galaxy idea was still hard
to take in general.}

In the third place,
the name ``black hole'' with all its psychological
and metaphorical implications 
(e.g., a {\it hole} lets you move somewhere else--perhaps into new physics)
did not exist.\footnote{Wheeler thrust the term ``black hole'' in 1967. See 
Wheeler, J.~A.~(1968), ``Our universe: The known and the unknown,'' 
{\it American Scholar, 37}, 248--274. It is important to note also that
the term had appeared in print as early as 1964. 
See
Ewing, A. (1964), ```Black holes' in space,'' {\it Science News Letter, 85}, 39. 
} 
Instead, the literature would talk about
``frozen stars,'' a quite anticlimactic term. 


\hfil

\noindent{\bf The ``Too Esoteric'' Argument}

\noindent
{\it General relativity was the ``string theory'' of the 1930s}


\hfil

``Very odd'' is how Oppenheimer described in writing
his new results to George Uhlenbeck.\footnote{Cassidy (ref.~\ref{cassidy}), 176.} 
This is probably an
understatement.







Three reasons made this odd situation even odder.
In the first place,
astronomy in general was much more distant from physics than it is today.
It had a natural history ring to it. Oppenheimer
was working on the margins of physics.


Secondly, the influential Arthur Eddington 
had given an esoteric twist to astronomy and cosmology,
invoking arguments that at times were perceived as too philosophical.

Thirdly, and most importantly,
the abstruse character of general relativity did not help either.
As late as 1960, Alfred Schild said that
``Einstein's theory of gravitation ... is moving from the realm of 
mathematics to that of physics''\footnote{Quoted in 
Kragh (ref.~\ref{kragh}), 362.}
and
even as late as 1958, Wheeler 
(the eventual champion of black holes)
did not feel
comfortable at all with the concept of black holes 
(which led to the famous Brussels confrontation with 
Oppenheimer that year).\footnote{See footnote~(\ref{thorne}).}

This situation became worse in the United States, as
Cassidy says,
where theoretical research was supposed to aid experimentalists, not
become involved in radical, creative, German-style 
speculations.\footnote{Cassidy (ref.~\ref{cassidy}), 179.}

On top of everything, 
one has to add the fact that Oppenheimer
worked with idealized spherical symmetry conditions in his 
treatment of black holes. Even though this approach 
is not considered particularly grave
today, back then spherical symmetry was considered a special,
probably physically irrelevant
case.\footnote{We are
grateful to Prof.~Robert Wagoner (Stanford) for his comment on this 
particular issue and for his interest in this paper's 
discussion. He mentioned that a similar opinion 
of special-case irrelevance surfaced with Kerr's solution for
black holes. Prof.~Robert Wald (University of Chicago), to whom we
are also grateful, offered further the case of Big Bang cosmology
as an example along these lines of special cases.}

One also 
has to keep in mind the precedent of Eddington's bashing 
of 
Chandrasekhar's ideas (referred to as ``stellar buffoonery'') on the collapse
of white dwarfs in 
1935.\footnote{Israel 
(ref.~\ref{300y}), 217.}
One may wonder just how influential this case 
might have been
as Oppenheimer was trying to expound his position.

\hfil

\noindent{\bf The ``Not Earned the Right'' Argument}

\noindent
{\it Intellectual seniority matters}

\hfil

We could phrase this argument thus:
if you had already succeeded at prestigious physics 
(which in that time meant something nuclear or particle),
as in the case of Bethe,\footnote{Bethe, H., 
\& Fermi, E.~(1932), ``\"Uber die Wechselwirkung von Zwei 
Elektronen,'' {\it Zeitschrift f\"ur Physik, 77}, 296--306.}
then you earned the right to do something
unorthodox in the border of physics and be taken seriously.


Since Oppenheimer liked to be at the center of things,
and was (intellectually) moving all the time,
he had never quite achieved fame in anything before
publishing his paper with Snyder
(Oppenheimer's papers with Max Born in 
1927\footnote{Born, M., \& Oppenheimer, R.~(1927), ``Zur Quantentheorie 
der Molekeln,'' {\it Annalen der Physik, 389}, 457--484.} 
and Melba Phillips in 1935\footnote{Oppenheimer, J. R., \& Phillips, M. 
(1935), ``Note on the transmutation function for deuterons,'' 
{\it Physical Review 48}, 500--502.} 
had presumably been his most famous, but these papers, 
with only 15 citations each\footnote{Google Scholar Citations.}  
in their respective first ten years,
could not be called truly revolutionary). 
This was
made worse by his distancing from physics into 
philosophy, literature and left-wing politics,
so 
this trend of going to extreme stellar physics
could be seen as part of a movement away from
mainstream physics. 

\hfil







\noindent{\bf The ``Wrong Episteme'' Argument}

\noindent
{\it The scientific world had already 
enough infinities to deal with}

\hfil

Why is it that
Einstein never accepted the black hole consequences of his theory?
One possibility is that black holes did not belong to the
correct episteme of the time.

Michel Foucault uses the term ``episteme'' to refer
to the implicit assumptions about how we know the 
world.\footnote{Foucault, M.~(1970), {\it The order of things}, New York:
Random/Vintage.}
More precisely, it refers to
``...the assumptions about knowledge, method, and theory
which at any given time period are shared across ``discursive formations''
(which as a first approximation can be translated as 
``disciplines'').''\footnote{Hess, D.~J.~(1995), {\it Science and technology
in a multicultural world}, New York: Columbia University Press,
87.}
An episteme differs from a Kuhnian paradigm in part
in that it is transdisciplinary.

These statements are best explained by examples.
According to Foucault's ideas, not just physics but 
the whole realm of academic knowledge in the beginning of the
twentieth century was marked by the episteme of equilibrium
and closedness.
One sees it in 
biology (population equilibrium theory),
economics (classical, pre-Keynesian theory),
linguistics (syntax rather than evolution),
the social sciences (structuralism),
and physics, as in Bohr's atom.\footnote{Ibid., 94.}




To these examples discussed by anthropologist of science David Hess, one may add
how Einstein develops his general theory of relativity
immersed in this episteme.
Einstein's model of the universe needs (by Einstein's own
later account) an artificial  
``cosmological term'' in order to preserve the equilibrium 
episteme.

Oppenheimer's stellar 
``indefinite contraction'' did not belong to this
episteme. Was he ahead of his time, sensing the forthcoming episteme
of open processes?

It is revealing that Einstein published a 
paper\footnote{Einstein (ref.~\ref{einstein}).}
with the intention of killing the Schwarzschild 
singularity\footnote{In 1916, Karl Schwarzschild found a solution
of Einstein's equations which were not well behaved at certain 
points. See
Schwarzschild, K.~(1916), 
``\"Uber das Gravitationsfeld eines Massenpunktes nach der Einsteinschen 
Theorie,'' 
{\it Sitzungsberichte K\"oniglich Preus}, Akad.~Wiss.~Berlin, Phys.-Math.~Klasse, 189--196.}
once and for all.
The paper was entitled
``On a stationary system with spherical symmetry
consisting of many gravitating masses.''
It used a {\it stationary} argument to show that
black holes were impossible.
What he actually proved was only that there are no stable
solutions to Schwarzschild radius stars (and therefore
his original intention was frustrated), but 
for some reason Einstein thought this proof was sufficient.
A case could be thus made for his tacit commitment to the
equilibrium episteme.

Along 
similar equilibrium-episteme lines, Eddington\footnote{Israel 
(ref.~\ref{300y}), 219.}

\begin{displayquote}
...like virtually every relativist of the time, 
considered the Schwarz- schild radius to be both a singularity
and an {\it impassible} barrier. The image that he
conjures up of the star {\it `at last finding peace'} is of a body
frozen at the Schwarzschild radius... [emphasis added]
\end{displayquote}

We might speculate 
what would an out-of-the-equilibrium-episteme
attitude
look like for a person living within the equilibrium episteme.
Perhaps an ``equilibrium epistemist'' would simply consider a person like
Oppenheimer as somewhat lost, not confident, confused.
The following 1967 quotation from particle physicist Isidor Rabi
(born in 1898, and therefore only six years Oppenheimer's senior) is 
useful:\footnote{Quoted in Thorne 
(ref.~\ref{thorne}), 208.}

\begin{displayquote}

[I]t seems to me that in some respects Oppenheimer was {\it overeducated}
in those fields which lie {\it outside} the scientific tradition,
such as his interest in religion, in the Hindu religion in particular, 
which resulted in a feeling for the mystery of the Universe that surrounded 
him almost
like a fog. He saw physics clearly, looking toward what had already been done,
but at the border {\it he tended to feel that there was much more of the
mysterious and novel than there actually was}. He was insufficiently confident
of the power of the intellectual tools he already possessed and did not drive his 
thought to the very end because {\it he felt instinctively that new ideas and new
methods were necessary to go further} than he and his students had already gone
[emphasis added].
\end{displayquote}

Rabi thus felt the need to explain Oppenheimer's
unassertiveness as something having nothing to do with science,
but rather with his other, extra-scientific inclinations.

It is appropriate to 
finish this section with an intriguing 
comment.\footnote{We are in debt to 
Prof.~Robert Wald (University of Chicago) and Randall Espinoza
(University of Illinois at Chicago) 
for this particular point.}
A different reading of Wheeler's initial 
attitude towards Oppenheimer could be made
(and it is one that does not necessarily contradict this article's main argument) 
in which it was Wheeler's strong commitment to a particle physics
point view which would have intensified his lack of interest in Oppenheimer's
work. 
We are referring in particular to Wheeler's work
on 
geometrodynamics\footnote{Wheeler, J. A. (1957), ``On the nature of quantum geometrodynamics,'' {\it Annals of Physics, 2}, 604--614.} 
and his aversion
to singularities, and on how gravity, considered as part of the particle physics
puzzle, could have helped to solve fundamental difficulties in the theory. 
In such a reading, Wheeler's disregard of Oppenheimer's ideas
would be less dramatic and more of a pragmatical nature. 
The details of such a study are to be carried out elsewhere.









\hfil

\noindent{\bf The ``Wrong Relativistic Ontology'' Argument}

\noindent
{\it The spell of geometry}

\hfil

The set of ten equations of general relativity,
\[
  G_{ab} = 8\pi \, T_{ab} \, ,
\]
can be interpreted in different ways.
If one reads them from right to left,
then the matter (through the momentum-energy tensor $T_{ab}$)
determines the geometry (described by Einstein tensor $G_{ab}$).
If, on the other hand, one chooses to read them 
from left to right, then geometry would be ontologically primal:
geometry dictates how matter must behave.

Even though in either interpretation one must
have of course exactly the same equations,
from a cognitive point of view, 
{\it and even from a mathematical point of view},
it makes a huge difference what interpretation
you adhere to.

The original interpretation was the geometrical one,
even to the point that
Einstein's crafting of general relativity is
imbued with quite a bit of implicit space-time reification,\footnote{Janssen, 
M.~(2007), What did Einstein know and when did he know it?,
in J. Renn (Ed.), {\it The Genesis of General Relativity}, vol.~2 (pp.~785--837),
Dordrecht: Springer, 825.} called ``substantivalism'' in the literature,
a curious state of affairs indeed since Einstein was an enemy of absolute space.

In any event, nowadays general relativity applications
follow a more ``matter first'' approach.
In this sense, Oppenheimer appears to be again ahead of his time. 
The crucial point is that the geometrical approach
biases your understanding 
and your problem searching towards more static/stationary situations.

To make this point clearer, consider a system of two masses
rapidly rotating around each other. This system will produce
oscillating space-time ripples moving away from them.
If one starts from the two masses, then it is straightforward
to calculate the surrounding oscillating geometry.
However, the opposite problem of reconstructing the 
masses' movements from the geometrical ripples is a fantastically
complicated problem. This is an example of a problem that does
not lend itself to be {\it formulated} if one starts from a
geometrical viewpoint.


The consequence of all this is that
your aesthetical judgment (``geometry first'') is going to have
an effect on the type of problems you tackle.
If you unite this effect with the equilibrium episteme 
one (described in the previous section) the result is devastating
for Oppenheimer, as collapsing stars are thus doubly denaturalized:
they are not in equilibrium, and they are not ``geometry first.''






Oppenheimer was an outsider to this geometrical ontology.
As Cassidy says, 
``the few active general relativity theorists were interested not in the
astrophysics of a star collapsing into a 
{\it mathematically awkward} singularity but in the more
{\it elegant and well-behaved geometry} of continuous, nonsingular
curved space-time'' (emphasis added).\footnote{Cassidy (ref.~\ref{cassidy}), 177.}


\hfil

\noindent{\bf Discussion: 
Oppenheimer's Black Hole vis-\`a-vis Einstein's EPR Paradox and
Zwicky's Dark Matter}

It is helpful to perform a comparison between the black hole idea 
as developed by Oppenheimer's group (in 1939) with the 
Einstein-Podolsky-Rosen Paradox (in 1935) and Zwicky's concept of dark matter 
or {\it Dunkle Materie} (in 1933).\footnote{This discussion actually originated
in lively fashion during the January 2014 Stanford 
meeting. Einstein, A., Podolsky, B., \& Rosen, N.~(1935), ``Can 
quantum-mechanical description of physical reality be considered 
complete?,'' {\it Physical Review, 47}, 777;
Zwicky, F.~(1933), ``Die Rotverschiebung von extragalaktischen Nebeln,'' 
{\it Helvetica Physica Acta, 6}, 110--127.}
(For the benefit of readers not familiar with these scientific concepts, 
there is a brief description of each in Appendix B.)

All three theoretical concepts appeared in the 1930s.
They all have in common that the related ideas were put aside
for several decades before they were taken seriously, when one could
say experiments made them inevitable.

The similarities stop there, though. 
The EPR Paradox was really not a discovery of a new entity,
but rather a gedankenexperiment designed with the sole
purpose of pointing out an inconsistency in the looming (for Einstein) 
conceptual edifice of
quantum mechanics.
Einstein would have been happy if he had caused the dismissal of
the Copenhagen interpretation of 
quantum mechanics; there was no actual intent, or interest, of carrying 
out the experiment or having somebody else carrying it out.

Zwicky's dark matter was also more about pointing out an inconsistency
than discovering a new substance. The fact that we are currently,
80 years later, 
looking for dark matter should not distract us from this point.

Both Zwicky's dark matter and Einstein's EPR Paradox are more
what one would call ``anomalies'' in the Kuhnian sense
(as is the case of Mercury's perihelion precession) than
true proposals/discoveries of new physical entities or phenomena.
This makes a huge difference, since anomalies tend to be 
treated with respect, and kept along
in their unresolvedness.

This is, we believe, what makes the history of Oppenheimer's black holes
much more intriguing than Einstein's and Zwicky's counterparts.

\hfil

\noindent{\bf Final Words}

Something rather interesting happened in physics in the late 1930s.
In what was to prove (judged in retrospect) as 
his last shot at intellectual glory, Oppenheimer,
with what might be termed the tacit complicity of the whole physics community, 
missed a chance to fully discover black holes---not observationally, 
but theoretically.
Says 
Werner Israel about Oppenheimer's work with Snyder: ``[it] has strong
claims to be considered the most daring and uncannily prophetic paper ever
published in the field.''\footnote{Israel 
(ref.~\ref{300y}), 226.}
Thorne says: ``This line of reasoning [what happens when a neutron star
cannot hold its own weight] is so obvious in retrospect that it seems
amazing that Zwicky did not pursue it, Chandrasekhar did not pursue it, Eddington 
did not pursue it.''\footnote{Thorne 
(ref.~\ref{thorne}), 178.}
We have to add that once it was initially pursued, by Oppenheimer, it
was then ignored
by the community until the late 1950s, many years after the war was over.

In this paper we have tried to address the issue of why is it that
this new idea did not receive the benefit of the doubt
in the same sense that other oddities did (such as many in particle physics,
e.g., the uncertainty principle),
even though there
is plenty of evidence that Oppenheimer's work was considered 
authoritative by at least some of his contemporaries, 
such as Hayashi in 1962 (in addition to what was discussed about Landau
above).

In this paper we have tried to go beyond previous
discussions on this topic.
The way we did so was by adding an additional layer
of a more history-of-ideas, Foucaldian nature.
We entertained the possibility
that at least some of the explanations might have to
do with 
idiosyncratic aspects (of the scientific culture, that is),
in particular to a tacit commitment to the
equilibrium episteme and a geometry-first ontology.
In contrast with the EPR Paradox and the ingenuity of the
dark matter concept, the black hole idea is not so 
much about pointing out an anomaly as adding a new object
to our universe.



In this essay, we have sought to supplement the existent literature 
on the subject by enriching the explanations 
and possibly complicating the guiding questions.
We suggest a rereading of Oppenheimer as a more intriguing,
human, ahead-of-his-time figure.



















\hfil

\hfil

\noindent{\bf Acknowledgements}

This work was supported by
grant 805-A4-125 of the Universidad de Costa Rica’s Vicerrector\1a 
de Investigaci\'on
and the CIGEFI, and by grant FI-0204-2012 of the MICITT and the
CONICIT.

\newpage

\noindent{\sc appendix a}


\noindent
{\bf Brief Description of the Relevant Papers}

\hfil

\noindent
{\bf Oppenheimer \& Serber, October 1, 1938} 

\noindent
{\it On the stability of stellar neutron cores}

This one-page, no-formula letter is a critique of Landau's work on
``condensed neutron cores''---as it was believed back then
that a neutron star (a ``neutron core'') could lie in the interior of stars 
like the Sun.\footnote{Landau's ideas appear on the following
two papers: 
Landau, L.~(1932), ``On the theory of stars,'' 
{\it Phys.~Z.~Sowjetunion, 1}, 285;
Landau, L.~(1938), ``Origin of stellar energy,'' {\it Nature, 141}, 333--334.}
 
The main point raised by the authors had to do with the necessary
inclusion of strong nuclear forces considerations
(which were absent in Landau's papers).
This inclusion was problematic because it came
in a moment in history in which
there was
``no existing nuclear experiment or theory [giving] a complete
answer to this question.''\footnote{Oppenheimer \& Serber
(ref.~\ref{oppserber}), 540.}

\hfil

\noindent
{\bf Oppenheimer \& Volkoff, February 15, 1939 (received January 3)}

\noindent 
{\it On massive neutron cores}

Here the authors continue commenting on 
improvements on Landau's ideas,
this time emphasizing the importance of using 
a general relativistic approach rather than a Newtonian one.
The reason for this is that neutron cores have an extremely high
density and require thus a 
relativistic approach.
Stars that would be stable in a Newtonian world are
unstable once general relativity is considered.

For the first time, the {\it indefinite contraction} fate for
heavy enough stars in mentioned.

\hfil

\noindent
{\bf Oppenheimer \& Snyder, September 1, 1939 (received July 10)} 

\noindent
{\it On continued gravitational contraction}

In this paper, the authors apply the equations of general 
relativity to prove that, at least under some simplifying 
conditions (non-rotating star, no pressure, no outward radiation), 
a large enough star will contract indefinitely.

This is the debut of black holes.
The authors describe how time
freezes at the Schwarzschild radius (of a few kilometers),
while it does not freeze for an infalling observer.

\newpage

\noindent
{\bf Einstein, October 1939 (received May 10)} 

\noindent
{\it On a stationary system with spherical symmetry
consisting of many gravitating masses}

After criticizing simpler treatments on the subject, 
Einstein uses a stationary argument (cluster of particles
in circular paths)
to argue that
black holes are impossible. However, what
he actually proves is that very compact stars are unstable.

\hfil


\noindent{\sc appendix b}

\noindent 
{\bf Brief Description of 
the 
EPR Paradox and Zwicky's Dark Matter}

\hfil

\noindent
{\bf Einstein-Podolsky-Rosen Paradox}

The EPR Paradox is a thought experiment designed to show that
there is a theoretical inconsistency within quantum mechanics
if one holds that it is a complete theory.
Imagine a pair of particles originating from a common source.
According to the Copenhagen interpretation of quantum mechanics, under some conditions
the state of particles 1 and 2
remain fundamentally undetermined until one decides to measure one
of them. When one does measure one of them, say particle 1, then 
either
$a)$ particle 2 has a definite state which, however, is not included
in the theory, rendering thus the theory {\it incomplete}, or
$b)$ particle 2 acquires, {\it immediately} after performing the
measurement on particle 1, certain definite physical property,
thus provoking an action-at-a-distance effect, which is contrary to the
principles of special relativity.

Einstein et al.~assumed that option $b)$ is untenable and thus quantum mechanics
must be incomplete---ruining thus the Copenhagen interpretation of the theory.
Experiments performed from the 1980s on have, however, corroborated option
$b)$.

\hfil

\noindent
{\bf Zwicky's Concept of Dark Matter}

The concept of ``dark matter'' was postulated in order to solve
a breach between theory and observation in astrophysics. As
stars move around the center of galaxies (including our own),
their speeds are higher than expected, 
as if there were a substantial amount 
of matter not accounted for: invisible matter---hence the name ``dark.''
More precisely, theory requires that
85\% (by modern calculations) of the mass be in the form
of dark matter. 
Or else, there is something fundamentally wrong with our theories of 
gravity. The dark matter problem has not been solved yet.

\newpage

\noindent{\sc appendix c} 


\noindent
{\bf Timeline of Events}

\noindent
Main events relevant to the discussion. Note the gap between 1939 and 1957.

\begin{table}[htp]
{\footnotesize{
\begin{center}
\begin{tabular}{ll}
1930s  	&heyday of nuclear physics, not so much of astrophysics and cosmology (which did not exist) \\ 
1930s  	&stellar energy production problem solved by Bethe\\
1932   	&discovery of the neutron\\
           	&Zwicky, Landau ask:  Are there neutron stars (or neutron cores inside stars)?\\
1938   	&JRO \& Serber:  do not forget to include nuclear forces, Landau\\
1939   	&JRO \& Volkoff:  do not forget to include general relativity\\
1939   	&JRO \& Snyder:  a large enough star will contract indefinitely\\
		&(at least under some simplifying assumptions)\\
1939   	&Einstein tries to show that black holes are not feasible, but what he actually \\
		&proves is that very compact objects are unstable\\ 
1939         &Landau allegedly adds JRO \& Snyder paper in his Golden List of classic         papers  \\
1957   	&Wheeler invents wormholes to explain away black holes; JRO not referenced \\
	         &the term ``wormhole'' appears thus many years before the term 
                   ``black hole'' \\
1950s 	&In the late 1950s, Wheeler and (independently) Zel'dovich theoretically    		consider black holes, \\
		& using computers, but neither of them publishes anything\\
1958  	&Finkelstein establishes the ``no return'' character of horizon; JRO
          not referenced \\
1958   	&Brussels confrontation between JRO and Wheeler, which is of a rather conceptual kind\\
1958    &JRO's work appears referenced by Landau in the English edition of
          Statistical Physics \\
1960   	&Kruskal paper (which is actually written by Wheeler) finally acknowledges black holes;\\
		&JRO not referenced\\
1962   	&authoritative review by Hayashi et al. on stellar physics references JRO's work\\
1963		&Wheeler becomes a supporter of black holes, lectures on them at the First Texas Symposium\\
		& (an international astrophysics conference) on December 1963, but JRO had lost all interest on\\
		&  the subject and does not even attend the talk even though he was present at the conference\\
1967 	&``black hole'' term made popular by Wheeler (even though it had been in print since 1964)\\
1967   	&JRO dies at 62 (on 18 February)\\
1967  	&first observation of pulsars (on November) by Hewish \& Bell group at Cambridge\\
1968		&first observation of pulsars published (February)\\
		&this corresponded to first ever Nobel Prize given to astronomers, in 1974\\
1969   	&first observation of millisecond pulsars (published in February), making plausible\\
		& the existence of compact objects such as black holes\\
\end{tabular}
\end{center}
\label{default}
}}
\end{table}%

\end{document}